# Collective information processing and pattern formation in swarms, flocks and crowds

Mehdi Moussaid[1, 2], Simon Garnier[2], Guy Theraulaz[2], and Dirk Helbing[1]

[1]ETH Zurich, Swiss Federal Institute of Technology, Chair of Sociology, UNO D11, Universitätstrasse 41, 8092 Zurich, Switzerland. Phone +41 44 632 88 81; Fax +41 44 632 17 67.

[2]Centre de Recherches sur la Cognition Animale, UMR-CNRS 5169, Université Paul Sabatier, Bât 4R3, 118 Route de Narbonne, 31062 Toulouse cedex 9, France. Phone: +33 5 61 55 64 41 ; Fax: +33 5 61 55 61 54.

**Keywords**: Self-organization - Social interactions - Information transfer - Living beings - Distributed cognition - Collective behaviors.

**Abstract**

The spontaneous organization of collective activities in animal groups and societies has attracted a considerable amount of attention over the last decade. This kind of coordination often permits group-living species to achieve collective tasks that are far beyond single individuals capabilities. In particular, a key benefit lies in the integration of partial knowledge of the environment at the collective level.

In this contribution we discuss various self-organization phenomena in animal swarms and human crowds from the point of view of information exchange among individuals. In particular, we provide a general description of collective dynamics across species and introduce a classification of these dynamics not only with respect to the way information is transferred among individuals, but also with regard to the knowledge processing at the collective level. Finally, we highlight the fact that the individuals' ability to learn from past experiences can have a feedback effect on the collective dynamics, as experienced with the development of behavioral conventions in pedestrian crowds.

# 1. Introduction

In nature, many group-living species - such as social arthropods, fish or humans - display collective order in space and time (see figure 1). In fish schools, for instance, the motion of each single fish is perfectly integrated into the group, so that the school often appears to move as a single coherent entity. In response to external perturbations, the whole school may suddenly change the swimming pattern, adopt a new configuration, or simply switch its direction of motion in near perfect unison. In case of predator attack, fish flee almost simultaneously, seemingly all aware of the danger at the same moment (see e.g. Partridge, 1982).

Similar coordinated collective behaviors can be found in humans (Helbing et al., 2001). Flows of people moving in opposite directions in a street, spontaneously organize in lanes of uniform walking direction, in this way enhancing the overall traffic efficiency by reducing the number of avoidance maneuvers.

A major characteristic of this collective organization lies in the fact that it emerges without any external control. No particular individual supervises the activities nor broadcasts relevant information to all the others and no blueprint or schedule is followed. This non-supervised order holds a puzzling question: By what means do hundreds or even thousands of individuals manage to coordinate their activity to such an extent without referring to a centralized control system?

Answering this question comes down to establishing a link between two distinct levels of observation: on the one hand, seen from a "macroscopic" level, the group displays a surprisingly robust and coherent organization that often favors an efficient use of the environment. However, on the other hand, from the "microscopic" point of view of a given individual, the situation is perceived at a local scale: the pedestrians, like the fish, do not have a complete picture of the overall structure they create. They rather react according to partial information available in their local environment or provided by other nearby group members.

The nature of the link between the individual and the collective level is investigated in this article. More specifically, the problem of how local interactions among individuals yield efficient collective organizations is addressed by studying how information is transferred among individuals. Indeed, the contrast between the limited information owned by single individuals and the "global knowledge" that would be required to coordinate the group's activity is often remarkable.

The unexpected birth - or emergence - of new patterns out of interactions between numerous subunits was first established in physico-chemical systems (Nicolis & Prigogine, 1977). Since then, it has been many times demonstrated that spontaneous order can appear in such systems because of the non-linear interactions among chemicals. Because the order emerges without external control these non-linear phenomena were labeled as self-organized.

Self-organization mechanisms are not limited to physical or chemical systems. During the last 30 years, they have also been identified in various living systems, such as cellular structures (Shapiro, 1988; Ben-Jacob et al., 1994; see Karsenti, 2008 for a review), animal societies (Camazine et al., 2001; Couzin & Krause, 2003; Sumpter, 2006; Garnier et al., 2007) or human crowds (Helbing & Molnar, 1995; Ball, 2004). Comprehending them is among today's most interesting challenges: first, because they are responsible for a significant part of the organization of animal and human societies; and second, because they are often the source of problems, such as vehicular traffic jams (Helbing, 1998), the spread of diseases (Newman, 2002), or the clogging of people fleeing away from a danger (Helbing, 2000).

The present study focuses on such behaviors in living beings: humans, like pedestrians, customers or Internet users, and animals, like insect colonies, vertebrate schools or flocks. Despite wide differences among these systems (in terms of the number of units, size or cognitive abilities of the individuals), human and animal systems can exhibit similar collective outcomes, suggesting the presence of common underlying mechanisms. For instance, bidirectional flows of pedestrians get organized in lanes (Helbing & Molnar, 1995), as well as some species of ants or termites (Couzin & Franks, 2002; Jander & Daumer, 1974); an audience of people may collectively synchronize their clapping (Neda et al., 2000) as fireflies synchronize their flashing (Buck & Buck, 1976); many insect species build trail systems in their environment, and so do humans (Hölldobler & Wilson, 1990; Helbing et al., 1997). Moreover, we choose to consider humans and animal systems because, unlike molecules involved in physical or chemical self-organized systems, living beings exchange and process information of multiple kinds when interacting with each other. This information influences and often determines the living being's next actions. In addition, the collective integration of individual knowledge often allows the group to produce efficient behavioral responses to their environment. Thus, studying the way individuals respond to information and how this information spreads among them is a crucial step for understanding the organizational abilities of many group-living species.

The following sections of our contribution are organized as follow: First, we start with a description of the major principles behind the concept of self-organization. Then, in section 3, we review various self-organization phenomena occurring in animal or human populations. Most of the discussed systems have been previously studied in the literature, but the novelty of this paper is to integrate them in a common framework based on the information exchange among individuals. That means, we highlight the internal mechanisms that allow the group to integrate and process this knowledge and to accomplish various tasks, such as sorting items, optimizing activities or making collective decisions. Accordingly, section 4 presents a generalized view of the dynamics on the "microscopic" and "macroscopic" levels of description, and a classification of the collective outcomes.

## 2. Self-organized behavior in social living beings

Because our purpose is to investigate the features of self-organized behavior, our first concern is to properly define this term and to bring major principles underlying such phenomena into the picture. A self-organization process can be defined as the spontaneous emergence of large-scale structure out of local interactions between the system's subunits. Moreover, the rules specifying interactions among the system's components are executed using only local information, without reference to the "global" pattern (Bonabeau et al., 1997). The distributed organization implies that no internal or external agent is supervising the process and that the collective pattern is not explicitly coded at the individual level. Furthermore, the emerging properties of the system cannot simply be understood as the sum of individual contributions.

Self-organization is a key concept to understand the relationship between local inter-individual interactions and collective patterns. A self-organized process relies on four basic elements:

1. A positive feedback loop, which makes the system respond to a perturbation by reinforcing this perturbation. Therefore, positive feedback often leads to explosive amplification, which promotes the creation of new structures. Typically, if the probability for an individual to perform a given action is somehow increased by other individuals in the neighborhood already performing the same action, the group is very likely to display a positive feedback loop. As an illustration, let us refer to a well-known experiment performed by Stanley Milgram in the streets of New York (Milgram et al., 1969): Milgram noticed that, when someone seems to look at something interesting in a particular direction, people around him tend to look in the same direction. More detailed studies showed that the tendency to imitate this behavior is approximately proportional to the number of surrounding people already looking in the same direction: a single person looking at a given point triggers 40% of naive by-passers to follow his or her gaze. This percentage grows to 80% and up to 90% with five and fifteen persons, respectively looking into the same direction. A positive feedback loop is in play: the higher the number of people looking in a given direction - let's say up in the air- the more likely surrounding walkers will look up in turn, increasing again the attractiveness of the looking-up behavior and so forth. This reinforcement dynamics usually leads to a non-linear propagation of a given behavior in the population.

2. The non-linear amplification of this snowball effect could eventually lead a system into a destructive state. Therefore, in self-organized systems, negative feedback typically sets in at larger perturbation amplitudes. Negative feedback dynamics are any kind of limiting factors that counteract the amplification loop, eventually leading to the stabilization of the collective pattern. These could be inhibitory or repulsive effects, but not necessarily so. For instance, why did the previous experiment not make the whole city of New York look up? Simply because, after some time, people tend to lose interest in the eye gazes and continue their walking. Hence, a more or less significant group of people looking up will form and stabilize, depending on the quality and relevance of information provided.

3. Self-organizing processes also rely on the presence of fluctuations. Random fluctuations constitute the initial perturbations triggering growth by means of positive feedbacks. People walking straight ahead toward their destination would never discover any point of interest in their environment, and a collective looking-up behavior would never appear. Instead, a weak tendency to check out the neighborhood may catch the attention of a few walkers, triggering the amplification loop and spreading the information into their neighborhood.

The unpredictability of exact individual behavior may also be the origin of the great flexibility of the system. As individuals do not deterministically respond to a given stimulus, there is a chance to discover alternative sources of information and other ways to solve a problem. In such a case, a positive feedback effect allows the system to leave a given state in favor of a better one.

4. Finally, self-organizing processes require multiple direct or indirect interactions among individuals to produce a higher-level, aggregate outcome. Repeated interactions among group members are the heart of any self-organized dynamics. Direct interactions imply some kind of direct communication between individuals (like visual or acoustic signals or physical contacts), while indirect interactions imply a physical modification of the environment that can be sensed later by other individuals. New York's by-passers unintentionally exchange information by means of direct interactions, namely by the visual signal they transmit when looking toward a particular direction.

On the basis of these four ingredients, it has been possible to describe and explain numerous collective behaviors observed in social insects and animal societies (Camazine et al., 2001, Couzin & Krause, 2003). Therefore, the concept of self-organization helps to elucidate the non-intuitive relationship between the apparent behavioral simplicity of group members and the complexity of the collective outcomes that emerge from their interactions.

We will now look at various case studies involving self-organized behaviors both in humans and animals, and describe them by means of the mechanisms introduced above. In doing so, we emphasize the distinction between the individual and the collective levels of observation, to better understand the relationships between both levels. Finally, we choose to classify the described systems according to the nature of the information transferred between individuals (i.e. either direct or indirect), because this difference has some further implications when studying the collective information processing, as discussed in the last section.

# 3. Case studies

## 3.1 Indirect information transfer

Indirect communication between individuals (also called stigmergic communication) is a frequent property of biological systems with many interacting agents. It refers to interactions that are mediated by the environment, based on the ability of individuals to modify their environment and to respond to such changes in specific ways. Stigmergy was originally introduced by French biologist Pierre-Paul Grassé at the end of the fifties to account for the coordination of building behavior in termites (Grassé, 1959, see Theraulaz & Bonabeau, 1999 for a historical review). Indeed, group-living insects often lay chemical signals in their environment to mark a particular location like a food source or to inform other group members of a recent change like a new construction stage in nest building. Signals exchanged in this way can be of different kinds, for example chemical or physical alterations of the environment. These alterations can often extend the duration of a signal and, as the marking of a personal territory shows, the spatial range as well. In humans, the signals exchanged can also exist within a virtual environment. Indeed, interactions within communities of people that have lately flourished on the Internet often go along with virtual signals left in blogs or forums. An interesting and simple example of such indirect information exchange involving virtual signals can be studied at the interactive website called *digg.com*, which we will focus on now.

*Case 1: The Online Social Network digg.com*

*Digg.com* is a website over which people can discover and share contents found elsewhere on the web. It allows its users to submit news stories they find while they browse the Internet. Each new story can be read by other community members. If they find it interesting, they can add a 'digg' to it. A digg is a virtual signal associated to a given story that can be seen by other users. The more diggs a story received in a given period of time, the more it becomes visible to the visitors, because news stories are displayed according to their popularity. In contrast to news magazines, however, popularity is not decided by some central decision maker, like a webmaster or editorial board, but by an automated algorithm that reacts to the number of diggs. Hence, the way news stories are displayed is determined by the activities of the users, and the interaction of users is mediated by the environment of the website, which classifies the interaction as indirect.

The dynamics at digg.com is an interesting case of decentralized collective organization to study. It turns out that interesting stories are widely spread among the community members at the expense of old or non-interesting ones. Moreover, the resulting system dynamics may be viewed as sorting the stories according to their relevance: At a given moment of time, the greater the number of diggs a story has received, the more interesting it is for the community. In the following, we discuss the underlying mechanisms of such a collective dynamics.

As pointed out before, interactions between users take place by means of indirect communication. Each user is capable of leaving a trace (the digg) in a virtual common environment, characterized by a multitude of more or less interesting stories. The behavioral rules of a given user can be summarized as follow: each user initially moves almost randomly through the environment provided by the website. In a neutral environment (i.e. in the absence of digged stories), each user has an approximately equally weak probability to read a given news, according to his or her own liking and interests. If the user encounters a story he or she finds relevant, he or she may modify the environment and mark the story for the attention of other members of the community.

Since popular stories are presented in an attractive way and easily accessible, the probability for another user to read a given story increases with the number of diggs the story has received. Therefore, a positive feedback loop can be identified here: the more a story is popular (that is to say considered relevant by users), the more likely it is to be paid attention to and to further increase its popularity. Consequently, interesting information is spread over the group in a non-linear way and the level of propagation of relevant stories increases exponentially with time. But such an exploding dynamics itself would lead a few stories to be so attractive that the great majority of the available information would remain unexplored. As described in the previous section, a negative feedback is needed to limit such self-amplification. Wu and Huberman observed that the decay in novelty of news counteracts the further amplification of their popularity: the older a news, the less it captures the attention of people (Wu & Huberman, 2007). The limited cognitive capacity of users and the competition of popular stories with a steady flow of incoming news for attention causes people to turn their attention to other stories. Accordingly, popular stories receive decreasing consideration as time goes by, and are finally replaced by other ones (figure 2).

Interestingly, it has been shown that the pattern of propagation of a novel information and the subsequent decay of attention depend on many factors, such as the time of the day it has appeared or the story's topic. This implies that the resulting sorting of the stories is somehow linked to the global environment: stories related to current events propagate faster than others. In terms of self-organizing mechanisms, this can be expressed by the fact that individuals tend to modulate their 'digging' behavior, with respect to the media-related context. Environmental specificities can thus induce a weak bias in the behavior of the users that would potentially result in a major change of the collective outcome. This sensitivity of the system provides a great flexibility in achieving the sorting task: different communities of people would sort the body of information in different ways, according to their interests, background and cultural environment.

*Case 2: Trail formation in ants*

In the animal world, one of the best studied examples of indirect communication is probably the trail formation in ant colonies. Many species of ants have the ability to lay chemicals, called pheromones, in their environment (Hölldobler & Wilson, 1990). Pheromones are a typical chemical support for information exchange in insect societies and can be used for various purposes such as warning of a danger, mating communication, or indicating the location of a food source (Wyatt, 2003). In particular, ants can deposit pheromone trails to mark the route from their nest to a newly discovered food source and share this crucial information with the rest of the colony. One can easily observe such a foraging behavior by setting out a piece of sugar in the neighborhood of a nest. After some time, more foragers appear at the food source, and soon an important flow of ants sets in between the nest and the piece of sugar (see figure 1a). How does the colony manage to establish such a foraging trail?

The process starts when a single ant finds a food source during a phase of random exploration. After feeding, the ant returns to the nest and drops small amounts of pheromones at regular intervals on its way back. This incipient trail has an attractive influence on other nestmates. Thus, although unaware of the food source location, nearby ants tend to modulate their random exploration behavior toward a trail-following behavior and may find the food source in turn. The greater the pheromone concentration, the higher the probability of an ant to follow the trail. Each new recruited ant finding the source reacts in the same way, returning to the nest and reinforcing the chemical trail with its own pheromones. This establishes a positive feedback: the more ants are recruited, the more attractive the trail becomes, increasing again the number of ants engaged in the process, and so forth. This leads to an exponential increase of the number of ants on the trail. However, pheromones are highly volatile chemicals. Thus, the evaporation of the trail can counterbalance its increasing attractiveness, leading the system to a stable state in which a constant flow of ants moves over the trail. A negative feedback occurs by other factors as well: it may result from the limited number of available foragers, from a competition between trails, or from the depletion of the food source. In any case, the negative feedback acts against the reinforcement loop, and a balance between opposite effects helps the system to stabilize in a new state, leading to a constant flow of ants on the trail (figure 3).

This ability of ants to leave marks in their environment constitutes a powerful means for efficiently spreading novel information. Interestingly, the way in which knowledge is processed at the group level provides many other benefits to the colony. In particular, controlled experiments reproducing ants' trail formation in the laboratory revealed that ants also carry information about the quality of the food source. Indeed, the workers tend to modulate their trail-laying intensity as a function of the quality of the discovered food (Beckers et al., 1993). From this behavioral modulation follows the ability of the colony to concentrate its effort toward the most profitable options. For example, if two food sources are available, the trail toward the richest one will be initially slightly more concentrated in pheromones than the others, and thus will attract a few more foragers at the beginning. However, as the number of workers involved increases, the difference in pheromone concentration between the trails grows as well, since the reinforcement operates faster on the path leading to the richest source. The feedback is further reinforced by the evaporation of the pheromones so that, finally, the competition between rich and poor sources directs the colony activity toward the most profitable option. If the selected food source runs out, ants stop laying pheromones and the trail vanishes, allowing the exploitation of other food sources. Based on the same reinforcement mechanisms, ants also manage to select the shortest route among several possibilities to reach a given food source (Beckers et al., 1990).

In contrast to the mechanisms in play at Digg.com, ants do not *sort* the different foraging alternatives according to their preference, but the colony rather selects the *best* option and focuses its foraging activity on it, almost ignoring all the others. The collective choice is decentralized: individual ants make no comparison of the different alternatives. The efficiency of the collective activities lies in the integration of information owned by single ants at the colony level, driving the group toward a consensus for the best foraging strategy.

*Case 3: Trail formation in pedestrians*

Humans are also often generating trail systems when walking through open natural space. One may observe such patterns imprinted in grassy areas in parks or meadows (figure 1d). The trails are caused by people walking off the originally planned ways, little by little trampling down the vegetation under their feet. The so-formed trail networks usually exhibit smooth curvy intersections and do not necessarily follow the shortest path between entry and exit points. Recent research highlighted that these trail systems result from a typical self-organization process (Helbing et al., 1997; Goldstone and Roberts, 2006).

Unlike digg.com users, pedestrians do not cooperate to build an efficient trail system. They are simply goal-oriented agents, each having its own starting point and destination, but all pursuing the same aim: walking comfortably and avoiding detours as much as possible. However, each walker unintentionally prints his or her own "solution" on the environment and thereby "shares" it with the other pedestrians. Indirect communication among people is achieved by altering the ground via the walkers' footsteps. The subsequent walkers spontaneously reconcile their goal-oriented behavior with a preference for walking on previously used and more comfortable ground. The system, therefore, has a reinforcement mechanism: trails attract walkers that in turn improve the trails and increase their attractiveness. Over time, and by using trails frequently, the system evolves toward a *compromise* between various direct trails. This enhances the walking comfort at minimum average detours.

To illustrate and validate this dynamics, Helbing et al. have developed an individual-based model of trail formation (the active walker model) (Helbing et al., 1997). The model is based on two intuitive behavioral rules: in a plain environment, each walker simply moves directly toward his or her destination point. However, such a movement prints a slight trail on the ground. If a pedestrian perceives such a trail on his or her way, he or she feels attracted toward this trail with an intensity proportional to the trail's closeness and visibility. The so-called walker model is complemented by a dynamic model of the ground structure, which is modified by walking pedestrians (e.g. by trampling down vegetation or leaving footprints in snow). This alteration of the ground is limited by a maximum trail intensity, to take into account the effect of saturation. The ground structure also changes in time owing to the regeneration of vegetation, leading to the slow but permanent restoration of the environment. Simulations made with a steady stream of pedestrians, all coming from and going to a few destinations at the periphery, gave rise to the formation of trails similar to those observed in urban grassy areas. In particular, the model predictions match several aspects of experimental trail systems generated when many people moving in a virtual environment try to minimize their travel costs by taking advantage of the trails left by others (Goldstone and Roberts, 2006).

These studies support the idea that a self-organized dynamics is the origin of trail formation by humans. Therefore, there exist some fundamental analogies in the mechanisms underlying pedestrian and ant trails formation. People modify their environment by means of their footsteps and, at the same time, feel attracted by this modification. Incipient trails are reinforced by a positive feedback loop that finally gives rise to persistent patterns. Evaporating pheromones in ant trails play the same role as regenerating vegetation in pedestrian paths, by counterbalancing the previous amplification effect. Pedestrians also take advantage of the trails they produce. Without any overall view of their environment, people collectively find a good compromise in terms of short, but comfortable ways linking several entry and exit points.

**3.2 Direct Information Transfer**

Information transfer can also occur through *direct* interactions. In this case, no modification of the environment (either real or virtual) is needed. Individuals rather behave according to the actions of their neighbors. For this reason, direct interactions are usually quite limited in their range (where a neighborhood may be defined in a metric or topological way). The information exchanged can be of different kinds, ranging from visual signals to acoustic ones, or physical contacts. This kind of interaction is at the origin of various spatio-temporal coordinated behaviors. In the following, we examine the dynamic of coordinated movements in fish schools, the emergence of temporal coordination in a clapping audience and the emergence of spatial coordination such as the formation of lanes observed in some species of ants as well as pedestrians.

*Case 1: Fish Schools*

The coordinated motion of schools of thousands or even millions, of individuals, all moving cohesively as a single unit, constitutes an interesting case to study. Various group-living animal species exhibit this remarkable ability to move in highly coherent groups, such as bird flocks (May, 1979; Higdon & Corrsin, 1978) or fish schools (Shaw, 1962; Partridge, 1982). We choose to focus on the abilities of fish to coordinate their movements in groups, primarily because they have been well studied, both from an empirical and a theoretical point of view.

Fish schools possess particular group-level properties. The observation of numerous individuals, all moving in parallel in the same direction and suddenly switching direction, implies that all individuals have somehow acquired the same turning information at almost the same moment. In case of a predator attack for example, the few individuals that perceive the danger trigger a wave of fleeing reactions that rapidly spreads across the school. Another feature of fish schooling is the variety of movement patterns that can be adopted. Spatial structures like mills, balls or vacuoles are examples of observable emerging organizations, the scales of which always exceed the size of a single individual by far (Parrish et al., 2002; figure 1b). Considering the enormous number of individuals involved, a centralized organization is hard to conceive. The most likely explanation of these group behaviors is self-organization.

Early experimental studies demonstrated that fish apply two different means of interaction: vision, used to acquire information about the motion of other fish, and the so-called "lateral line system", a sense organ located along the side of the fish that responds to water movement, providing information about the distance of neighboring fish (Partridge & Pitcher, 1980). Individual-based models have been developed on the basis of these observations (Aoki, 1982; Huth & Wissel, 1992; see also Reynolds, 1987 for a very influential flocking algorithm). Huth and Wissel suggested that each fish within a school follows a set of simple rules to determine its next position according to the position and orientation of its closest neighbors. In its simplest form, the model proposes that each fish $i$, located at position $\vec{x}_i$, adjusts its direction vector $\vec{v}_i$ at each time step by turning an angle $\alpha_{ij}$, where $\alpha_{ij}$ depends on the distance $r_{ij} = |\vec{x}_j - \vec{x}_i|$ and the velocity $\vec{v}_j$ of other fish $j$ in the neighborhood. In particular, the model suggests that fish can adopt three distinct behaviors according to the spatial proximity of the neighbors:

1. At short distance, when $r_{ij} \leq r_1$, a fish $i$ shows a repulsive behavior to avoid a collision. Within this distance range, fish $i$ turns perpendicularly away from the swimming direction of fish $j$, leading to $\alpha_{ij} = \min(\langle \vec{v}_i, \vec{v}_j \rangle + 90°, \langle \vec{v}_i, \vec{v}_j \rangle - 90°)$, where $\langle \vec{v}_i, \vec{v}_j \rangle$ denotes the angle between the swimming directions $\vec{v}_i$ and $\vec{v}_j$.

2. At intermediate distances, when $r_1 < r_{ij} \leq r_2$, fish $i$ aligns itself with fish $j$. The related angle $\alpha_{ij}$ is thus defined as $\alpha_{ij} = \langle \vec{v}_i, \vec{v}_j \rangle$.

3. At large distances, when $r_2 < r_{ij} \leq r_3$, fish $i$ is attracted by fish $j$ to maintain cohesion within the fish school and turns according to $\alpha_{ij} = \langle \vec{v}_i, \vec{x}_j - \vec{x}_i \rangle$.

When fish are too far away to sense each other (i.e. $r_{ij} > r_3$), no interaction takes place between the individuals, and the direction vector $\vec{v}_i$ remains unchanged. Simultaneous interactions are determined by calculating the arithmetic average angle $\alpha_i = \frac{1}{k} \sum_{j \neq i}^{k} \alpha_{ij}$, where k is the number of interaction partners. Finally, imperfect sensing and responses of fish are taken into account by choosing the effective turning angle according to a normal distribution with mean $\alpha_i$ and standard deviation $\sigma$. To account for the limited information processing capacity of fish, the number of simultaneous interacting partners is restricted to the k nearest neighbors. Computational results show that the model generates coherent schools for k>3, while k>4 do not further improve the model performance (Huth & Wissel, 1992; Camazine et al., 2001). Therefore, the value k=4 is often chosen in the literature. Typical parameter values are $r_1 = 0.5L$, $r_2 = 2L$ and $r_3 = 5L$ (where L is the body length of a fish). Several improvements of the model such as the consideration of a "blind area" behind the fish or a higher weight of the avoidance behavior can be made to enhance the realism of the model. However, they were shown to have little influence on the collective behavior.

Simulations based on such simple behavioral rules generate convincing schooling with no need of any supervision. Sudden moves of fish are imitated by their close neighbors. The higher the number of fish adopting a given behavior, the faster this behavior propagates among previously uninformed individuals. This reinforcement process leads to a quickly increasing number of fish responding to new information. The negative feedback here is simply given by the limited number of individuals, which inhibits the previous amplification. Finally, the interplay between positive and negative feedbacks gives rise to an S-shaped dynamics as described for other systems (e.g. figure 2b and figure 3). That is, the sudden increase of the number of individuals adopting the new swimming direction is followed by a saturation effect.

Predictions of the above model have been compared with various experimental datasets (Huth & Wissel, 1994). The simulations results agree with experimental data in many points, such as the distribution of distances to the nearest neighbor, the polarization of the group, the average time a fish spends in front of the school, and many schooling patterns. This evidence allows one to conclude that the model captures the basic mechanisms underlying the phenomenon well. Interestingly, Huth and Wissel also demonstrate that changing the value of parameters r1 and r2 generates different group polarization levels, matching those observed in different species of fish. Similarly, Couzin et al. showed that these two parameters have a critical influence on the collective configuration the fish school adopts (Couzin et al., 2002, Gautrais et al., 2008). In particular, the study shows that changing the alignment range from small to large values results in the school forming packed swarms, mills (where individuals circle around their center of mass, Fig.1b) and parallel motion of the entire group into a common direction, respectively. This implies that individuals may adapt their interaction rules in a context-dependent way. In case of danger, stronger attraction and alignment make the group more sensitive to external perturbations and provide fast answers to external threats. In other contexts, however, weaker interactions can be more efficient, since the group does not systematically respond to each small fluctuation. Given a small alignment range, only the most relevant information is amplified, which allows the school to ignore stimuli of lower intensity.

*Case 2: Synchronized Clapping of an Audience*

Self-organizing mechanisms can also lead to the emergence of collective temporal coordination. The next case focuses on emerging synchronous activity that can be found in humans, when an audience showing its appreciation after a good performance suddenly turns from incoherent clapping into coordinated rhythmic applause. Although no particular rhythm is imposed by any supervisory control, a common clapping frequency and phase emerges from the interaction between people.

Audience members interact by means of the acoustic signal produced by each clap and heard by other audience members. In such a way, people communicate their clapping rhythm to their neighbors, and acquire information about the rhythm adopted by the others around.

Similarly to fish behavior in schools, people tend to adjust their activity with respect to the average information they get from their nearby environment. In the beginning, small clusters of synchronized individuals may appear by chance. This locally stronger information, then, produces a positive feedback loop: the more individuals locally agree on a clapping rhythm, the stronger is their influence on other audience members. This results in the spread and amplification of common rhythmic activity among the spectators, and the whole audience finally achieves a consensus on their clapping rhythm. This reinforcement process is widespread in other natural systems (Strogatz, 2003). On the basis of similar mechanisms, some species of fireflies can achieve flashing synchronization (Buck & Buck, 1976). However, a quantitative analysis of recordings of audiences in Eastern European theaters and concert halls revealed a major difference compared to other animal synchronous activities. Néda et al. (2000) identified a particular common pattern characterized by an initial phase of incoherent but loud clapping, followed by a transition to synchronized clapping, which was again replaced by unsynchronized applause, and so on (Figure 5). Such a dynamics has not been observed in fireflies for example, although the underlying mechanisms are similar (individuals are adjusting to the average rhythm of their neighbors).

In order to interpret this alternation of ordered and disordered states, the authors relied on a model of coupled oscillators, originally suggested by Kuramoto (Kuramoto, 1975). The model is well adapted to audience behavior and shows that a large number of oscillators coupled together (continually adjusting their frequency to be nearer to the average) will finally oscillate synchronously, provided that the distribution of initial frequencies of oscillators is not greater than a critical value (Kuramoto, 1984). As pointed out by the authors, however, this model does not explain the wave-like aspect of synchronized clapping: a large dispersion of the initial clapping frequency would not lead to any synchronized state, while a smaller one would produce a persistent rhythmic applause as in fireflies, but the alternation between the two regimes is not theoretically expected.

Interestingly, experimental observations of individual clapping behaviors reveal two possible modes of clapping: a loud and fast clapping mode, characterized by a large frequency distribution, and a slower one, characterized by a smaller dispersion of frequencies. An interpretation of the wave-like synchronization directly follows from these observations: the first mode is initially adopted by the audience and leads to a random applause regime, as expected by Kuramoto's model. Then, depending on the quality of the performance, the mood of the audience, or even cultural aspects of such behavior, a majority of the spectators may switch to the second clapping mode and give rise to coordinated applause. The resulting outcome is synchronized, but less noisy. The theoretical impossibility for an audience to combine loud and synchronized clapping leads to what the authors call the frustration of the system. Therefore, it may happen that the lower sound level which goes with coordinated clapping motivates enthusiastic audience members to clap louder, increase the frequency of clapping beyond a critical limit, where rhythmic coordination is possible, which causes an intermediate loss of collective coordination, until the slow mode re-establishes again.

The example shows how the emerging collective pattern can be sensitive to particularities of the group members' behavior. Compared to the coordination of fireflies exhibiting a continuous coordinated regime, people's behavior is subtler and the context of the situation influences the homogeneity of the clapping frequency, leading to the observed wave-like pattern.

Interestingly, in addition to the rhythmic information transferred among people, this example exhibits a second kind of information communicating the intention to start rhythmic applause. A sufficient amount of people switching to the second clapping mode propagates this intention of coordinated clapping to the rest of the audience and carries them along in a collective expression of enthusiasm. Similarly to fish schools that are capable of adjusting their behaviors in a context-dependent way, audience members modulate their clapping behavior to achieve a particular collective outcome. In humans however, the process appears to be highly cultural, as synchronous clapping appears very often in Eastern Europe, while the phenomenon is rare in North America.

*Case 3: Lane Formation in Ants*

We have previously seen and discussed how ant colonies manage to build pheromone trails, i.e. some sort of invisible highways between their nest and a relevant point of their environment (typically a food source). Throughout the description of the phenomenon we assumed that only indirect interactions between ants play a role. In certain species of ants however, the traffic over these trails may become so crowded that ants encounter frequent physical contacts and need to evade each other. In such a case, direct interactions also come into play as well. These are the origin of another emergent pattern called "lane formation". A similar phenomenon was observed in humans (Helbing, 1991).

As described in the previous section, many ant species create chemical trail networks for exploration, emigration or transportation of resources. The functioning of such a system strongly depends on an effective management of traffic along the trails. In the neotropical army ants *Eciton burchelli*, the flow of traffic along trails is known to be particularly important (Schneirla, 1971; Gotwald, 1996). Colonies of this species organize large hunting raids that may involve more than 200,000 individuals. The main foraging trail is composed of two flows of ants: one corresponding to individuals moving from their nest to the end of the trail and the other corresponding to ants carrying prey and returning to the nest. Observations show that the bidirectional traffic in army ants organizes into lanes (Franks, 1985): ants returning to the nest occupy the center of the trail, while ants leaving the nest predominantly use both margins of the trail, in this way protecting prey from enemies.

How do the lanes emerge in this system? First, as described in the previous section, a dense traffic is established along the trail by means of indirect interactions via pheromones. This can be observed in many other ant species, so it does not explain the emergence of lanes itself. In case of army ants, an additional mechanism based on direct interactions is responsible for the spatial structuring. A single ant can perceive other ants at short distance and tends to turn away from them within this short-range interaction zone. This kind of avoidance behavior can account for the formation of lanes in any kind of oppositely driven particles, as a simple result of physical interactions: individuals meeting others head on tend to move aside as a result of the repulsive effect. But as soon as they happen to move behind each other in the same direction, a more stable state has formed, in which side movements are no longer needed. The reinforcement of this incipient organization is based on the fact that the probability of an individual leaving an existing lane decreases as a function of the lane size. Therefore, a positive feedback loop supports the formation of lanes across the population. The theory predicts that the number and shape of lanes are functions of the available space, the in- and outflows, and the fluctuation level (Helbing and Molnar, 1995; Helbing and Vicsek, 1999). However, traffic in army ants exhibits a fixed three-lanes structure regardless of external parameters. The reason for this unexpected configuration lies in the characteristics of ant behavior. Experimental measurements of the turning rate of individual ants show a quantitative difference between the behavior of ants leaving the nest and those returning to it: the former exhibit a higher turning angle during avoidance maneuvers than the latter (Couzin & Franks, 2002). This difference in the individual behavior of ants can potentially be explained by the fact that most of the ants returning to the nest are burdened with prey: due to their greater inertia, their turning requires more effort than for unloaded ants leaving the nest. On the basis of these observations, a simple model of the movement of ants along a pheromone trail can account for the observed pattern of organization. Simulations show that the heterogeneity in ants' turning range is enough to make the system organize in three lanes: outbound ants moving along both margins of the trail and returning ants using the center (Figure 4, see Couzin & Franks, 2002 for details of the model). Moreover, the exploration of the model parameters shows that this spatial configuration vanishes when the population becomes homogeneous, indicating that the value of the maximum turning angle has a critical influence on the emerging pattern.

Interestingly, the case of army ants demonstrates that, beyond the typical mechanism of lane formation, a simple behavioral specificity may result in significant characteristics of the collective pattern. Here, the difference between outbound and returning ants produces a slight asymmetry, when two ants of opposite flows interact. Although very weak, the bias gets reinforced, and individuals with a higher turning rate finally end up on the sides of the trail.

*Case 4: Lane Formation in Pedestrians*

Under everyday conditions, pedestrians walking in opposite directions also tend to organize in lanes of uniform walking direction (Milgram & Toch, 1969; figure 1c). In terms of traffic efficiency, this segregation phenomenon reduces the number of encounters with oppositely moving pedestrians and enhances the walking comfort. Here, people interact by means of visual cues. The information exchanged between walkers is somehow related to the most comfortable area to walk through in order to avoid unnecessary speed decreases and avoidance maneuvers. Indeed, a pedestrian within a crowd tends to adjust his or her normal goal-oriented behavior with respect to other people perceived in the neighborhood. Based on such simple assumptions regarding the behavior of walkers, individual-based models of pedestrian behavior have contributed to develop an understanding of the collective dynamics of people within a crowd. In particular, the so-called social force model (Helbing, 1991; Helbing and Molnar, 1995) was one of the first successful simulation models of self-organization in humans and has proved to be capable of capturing many complex patterns of motion, like the phenomena of lane formation, oscillations at bottlenecks and clogging effects (Helbing et al., 2005). The model describes the motion of a pedestrian $i$ at place $\vec{x}_i(t)$ by means of a vectorial quantity $\vec{F}_i$, reflecting his or her psychological motivation to move in a particular direction. Accordingly, the velocity $\vec{v}_i(t) = d\vec{x}_i/dt$ of pedestrian $i$ is given by the acceleration equation $d\vec{v}_i(t)/dt = \vec{F}_i(t) + \vec{\varepsilon}(t)$, where $\vec{\varepsilon}(t)$ is a fluctuation term that takes into account random variations of behavior. The acceleration force $\vec{F}_i(t)$ is the sum of several terms denoting different motivations of pedestrians. In the following, we present their simplest specification:

1. A driving force, $\vec{D}_i$, which lets the pedestrian $i$ move in his or her desired direction $\vec{e}_i$ at the desired speed $v_i^o$. The driving force is set such that the pedestrian adjusts the current velocity $\vec{v}_i$ to the desired one $v_i^o \vec{e}_i$, within a certain relaxation time $\tau_i$. This implies $\vec{D}_i = \dfrac{v_i^o \vec{e}_i - \vec{v}_i}{\tau_i}$.

2. A set of repulsive forces $\sum_{j \neq i} \vec{R}_{ij}$, which makes pedestrian $i$ avoid other pedestrians $j$ by moving away from them. In its simplest form, the term $\vec{R}_{ij}$ is defined as a gradient of a repulsion potential, resulting in $\vec{R}_{ij} = A_i e^{-d_{ij}/B_i} \vec{d}_{ij}$, where $\vec{d}_{ij}$ is the normalized vector pointing from $j$ to $i$, and $d_{ij}$ is the distance between the pedestrians; $A_i$ and $B_i$ are model parameters reflecting the strength and the range of the interaction, respectively.

3. A set of repulsive forces $\sum_k \vec{W}_{ik}$, which makes pedestrian $i$ to keep a certain distance from walls

and obstacles $k$. The influence of an obstacle $k$ is defined as a function of the distance $d_{ik}$ to the closest point of that obstacle: $\vec{W}_{ik} = A_k e^{-d_{ik}/B_k} \vec{d}_{ik}$, where $\vec{d}_{ik}$ is the normalized vector pointing from $k$ to pedestrian $i$, $A_k$ and $B_k$ are model parameters.

Further sources of influence can be added to the specification of $\vec{F}_i(t)$ as well, for example attractive forces modelling groups of people walking together or friction forces in very crowded situations. Recently, many studies make use of tracking algorithms to reconstruct trajectories of interacting pedestrians from video recordings taken in streets, train stations or highly crowded areas (Johansson et al., 2007, 2008). The analysis of such datasets allowed researchers to calibrate pedestrian models and to specify the interaction forces more precisely, based on a minimization of the error between observations and model predictions. Although this does not constitute a full validation of the underlying assumptions, the concept of social forces turns out to be versatile enough to account well for naturally occurring crowd patterns. This includes the formation of lanes in oppositely moving flows (figure 6), and unexpected transitions from laminar to stop-and-go and turbulent flows observed in areas of extreme densities (Yu & Johansson, 2007).

The previous case of lane formation in ants showed how some behavioral characteristics are very likely to shape the resulting pattern into a particular spatial configuration. Are there any similar features in the motion of pedestrians? In fact, people are often reported to have a preferred side of walking. In continental Europe for instance, lanes form more often on the right-hand side, regardless of the car-driving practices, while in Japan or Korea pedestrians are reported to walk on the left-hand side. Figure 1c, for example, shows asymmetrical lane formation in London, biased toward the right-hand side. Game-theoretical models suggest that an emerging behavioral convention could be at the origin of this asymmetric configuration (Helbing, 1991). According to this, it is more efficient to avoid someone on the side that is preferred by the majority. For such reasons, any random slight majority will cause further reinforcements, which ends up with a quite pronounced majority of people using the same avoidance strategy. This model implicitly assumes imitative strategy changes. One may also formulate this in terms of learning: Initially, pedestrians avoiding each other would have the same probability to choose the right or left-hand side. However, successful avoidance maneuvers would cause a more frequent use of the individual avoidance strategy. It turns out that such a reinforcement learning model eventually leads to an emergent asymmetry in the avoidance behavior, i.e. the probability to choose that side again on the subsequent interactions is increased. Simulations actually predict that different side preferences would emerge in different regions of the world, as observed (Helbing et al., 2001).

Two different levels of emergent behaviors are involved here at the same time. On short time scales, the way people avoid each other leads to the formation of lanes, which enhances the overall traffic efficiency. This phenomenon does not require any learning or memory about past interactions. In parallel, on longer time-scales, repeated interactions between pedestrians coupled to human learning abilities result in a further optimization of the traffic by establishing asymmetric avoidance behavior. This self-organization mechanism acts at the level of the population and induces a common bias in the people's behavior, which shapes the lanes into a particular configuration.

## 4. Discussion

### 4.1 General dynamics

In this paper we have considered various features of self-organization processes in human crowds and animal swarms. In all examples of collective behaviors, the description of the individuals' behavioral rules and the related feedback mechanisms allowed us to better grasp the underlying dynamics. In particular, the separate analysis of individual and collective levels of observation could highlight a common scheme of description of these systems. From the "microscopic" point of view, the behavior of a single individual can be characterized by providing answers to the following questions:

1. How does a single individual behave in the absence of information about the perceived environment?

2. What kind of information does it acquire in its neighborhood?

3. How does it respond to this information?

4. How is this information transferred to other group members?

Correspondingly, a model of the dynamics on the individual level can be constructed. First, each individual moves in its environment according to its spontaneous behavior. Here, we call spontaneous behavior the way in which group members move in the absence of new information regarding other individuals. For example, pedestrians usually have a spontaneous goal-oriented behavior. Without interactions, they simply move straight toward their next destination. Characteristics of this behavior are the speed of motion, the spontaneous probability of performing a given action, or environmental specificities that make the individual behave in a particular way.

At the same time, an individual may acquire information about its local neighborhood. This can happen by means of direct or indirect information transfer. As a result, the individual produces a behavioral response that stimulates or inhibits a particular behavior. This behavioral change is often proportional to the intensity or the quality of the acquired information. Finally, this adjustment results in a local spreading of the information. Once other individuals acquire the information, they adjust their behaviors in turn and propagate the information through the system. Table 1 summarizes the answers to the previous questions in the different examples discussed before.

From the local interactions between individuals, one can derive the aggregate dynamics of such systems, thereby connecting the "macroscopic" and "microscopic" levels of observation. In the beginning, the group often remains in a disorganized state, until a weak perturbation appears within the system. A perturbation is the occurrence of novel information within the group (like the discovery of a food source, a new digged story or a predator strike), or could also have a random origin. Then, depending on the size of the group and the nature of information exchange among the individuals, a positive feedback loop may be established: the number of individuals sharing the new information and modulating their behavior accordingly increases in a non-linear way. Typically, when an individual acquires the information "*There is something above*", it tends to look up, increasing the probability of other individuals to gain the information in turn and so forth. Eventually, negative feedback loops come into play (often induced by physical constraints like the limited number of individuals), and counterbalance the previous reinforcement. This helps to keep the amplification under control and yields a stabilization of a particular spatio-temporal pattern in the system.

## 4.2 Sensitivity to behavioral traits

On the basis of the discussed cases, two features of individual-level behaviors often induce significant changes at the collective level: the specificities of the spontaneous behavior of individuals and those of the behavioral response to new information (which correspond to the questions 1 and 3 above).

A key factor that may affect the spontaneous behavior of an individual is the presence of heterogeneity in its environment. The impact of such environmental specificities can turn out to be crucial, because a slight bias in individual behavior can be amplified through reinforcement loops and lead to major changes in the resulting pattern of behavior. For example, many animal species are strongly affected by the presence of physical heterogeneities in their environment (such as walls or edges). In fact, animals often search to maximize the amount of body area in contact with a solid surface, which provides protection against potential predators. This individual sensitivity to the environment has a strong influence on trail formation in ants: it has been demonstrated that the final shape of the trail formed between two points is strongly biased by the presence of a wall (Dussutour et al., 2005). Owing to an individual ant's tendency to move along a boundary, the positive feedback loop is likely to reinforce this bias and to be triggered faster in the neighborhood of a wall. Consequently, the resulting pattern is often unbalanced with respect to the wall's location. Likewise, temperature variation (Challet et al., 2005) or local air flows (Jost et al., 2007) can shape the outcome of the colony in a very different way. Similar environment-induced biases are likely to play an important role in the formation of trails in humans. In fact, according to the related model, the spatial distribution of the pedestrians' destination points directly determines the resulting trail network topology. In the same way, the presence of attractive or repulsive areas in the environment may shape the final trail system asymmetrically, even in the case of symmetrical origin-destination flows. Similarly, the influence of public media is likely to induce biases in the behavior of digg.com users. The initial probability to read a new story can, therefore, become affected, slightly favoring actual events and pushing this news to propagate faster across the community.

In the same manner, specificities of the behavioral response of group members to new information can create completely different emergent patterns. Several examples of this effect have been given in case of lane formation. Segregated lane patterns emerge both in bidirectional traffic of pedestrians and certain species of ants. The study of these phenomena showed that the number of emerging lanes in pedestrians is variable, depending on the density of people, the width of the street or heterogeneity in walking speeds. In ants, however, there is a fixed three-lane configuration (two lanes along the margin of the trail and one in the center, regardless of external parameters. The underlying segregation mechanism in ants and pedestrians are the same. However, in ants one of the two flow directions is restricted by heavy loads and, thus, cannot flexibly respond to interactions. The limited turning capabilities of such ants produce an asymmetry in the system and finally lead to the observed three-lane configuration. Such a phenomenon is conceivable in humans as well, for example in situations where heavily loaded pedestrians walk in one direction and unloaded one moves in the opposite direction (e.g. observable at railway stations). Similarly, we have underlined the fact that pedestrian lanes have a preferred side of the street. This could be interpreted as the result of a bias in pedestrian avoidance behavior during local interactions (Helbing, 1995). This illustrates, again, how a small change in the way individuals respond to interactions can lead to major qualitative differences in the resulting collective pattern.

## 4.3 Collective information processing

The above-described self-organization mechanisms constitute a powerful means by which a large number of individuals can achieve specific tasks that are often beyond the single individual's abilities, particularly when talking about animals. Although each group member acquires and spreads information locally, and this information is often limited and unreliable, the system as a whole fulfills higher-level tasks as if it had a global knowledge of the environment (Bonabeau et al., 1999). Among the cases described before, three kinds of collective outcomes can be identified: sorting, optimization and consensus formation.

**Sorting**: The dynamics underlying the website digg.com constitutes a typical example of a self-organized sorting procedure. The more relevant a story, the more often it is 'digged'. Therefore, the number of diggs a story gets attests for its rank at a given moment of time. The website thus acts as an information sorting system. The sorting is dynamic: the relevance of a given story is a subjective feature that depends on the users' interests, who choose to digg it or not. Consequently, according to the system's sensitivity to individual behaviors, the emerging classification of the stories is likely to vary between different communities, with respect to their cultural background, interests or goals. Various other self-organized systems generate such sorting of elements present in the environment. In some species of ants, for example, eggs are sorted out by workers according to their developmental stage and grouped into heaps of the same category. In this system, a positive feedback loop arises from the tendency of ants to deposit the egg they carry closer to a heap of elements of the same size (Deneubourg et al., 1991). In human populations, the segregation of people of different origins, social class or opinions follows a similar kind of non-linear dynamics and exhibits the main characteristics of a self-organized process (Schelling, 1969). In that case, the "sorting" process acts on the involved individuals themselves rather than on external elements of the environment.

**Reaching consensus:** Self-organized processes can also cause a group to reach a consensus. Achieving consensus on a given behavior is an essential aspect of collective organization, since it allows the individuals to act cohesively and prevents the group from splitting. Moreover, in most cases the consensus points toward the best alternative, which is often referred as "the wisdom of crowds" and based on an efficient collective integration of information (Surowiecki, 2004). In the case of foraging ants, the mechanisms underlying the recruitment of new workers leads the colony to choose among foraging strategies of different profitability. The presence of several alternatives (e.g. several food sources or several paths toward a given food source) systematically results in a common decision about which option the colony will concentrate its activity on. The solution that is amplified faster is usually chosen at the expense of the others. In particular, if a given solution provides a higher benefit to the colony (e.g. a richer food source), signal modulation favors information related to this option, and the entire colony finally focuses on it. Similarly, the large number of fish that constitutes a school reaches a collective consensus on the swimming direction. In particular, models show that the larger a school, the more it will be receptive to the information provided by a small percentage of informed individuals, which finally induce the schools to move toward a relevant destination (Couzin, 2005). The emergence of synchronized applause in an audience is another illustration, where numerous people achieve a consensus on their clapping rhythm.

**Optimization:** Finally, the third collective task highlighted by the case studies is the optimization of the group's activities. The formation of lanes in the bidirectional movements of ants and pedestrians is a form of traffic optimization. In both systems, repeated encounters with other individuals moving in the opposite direction constitute a serious disturbance of efficient and collective motion. The organization into lanes reduces the interaction frequency and the number of necessary braking or avoidance maneuvers. In such a way, the traffic efficiency is optimized. In humans, the additional emergence of walking *conventions,* such as a common preferred side of avoidance, further enhances the efficiency of traffic (Helbing et al., 2001). Likewise, the occurrence of trail systems allows pedestrians to optimize their travel from one point to another by finding a compromise that minimizes detours while maximizing the comfort of walking.

Throughout this paper, we differentiate direct and indirect information transfer. In the accomplishment of consensus, sorting and optimization tasks, both kinds of communication can be used. This implies questions regarding the specificities of the two communication methods in the execution of the different tasks. The examples of news sorting at digg.com, path selection in ants and trail formation in pedestrians illustrate the usage of **indirect information transfer** in the achievement of the different kinds of tasks. The prime specificity of indirect communication is that the collective solution to a given problem is mediated via the environment. Diggs popularity distribution, pedestrian trails and pheromone paths remain in the environment, sometimes even after the activity has ceased. Therefore, solutions emerging from indirect interactions are characterized by a high level of robustness to external perturbations. It is known, for example, that Pharaoh's ants make use of long lasting pheromones that remain attractive for several days to locate persistent food sources and ensure their exploitation from day to day, even when the foraging activity has to be temporarily interrupted (Jackson & Ratnieks, 2006). However, robustness to changes also implies lower flexibility. This shortcoming can be illustrated by the fact that, once an ant colony has selected a food source and built a trail toward it, it usually does not redirect its activity towards a better food source that appears at a later time, and stays stuck in a suboptimal solution (Pasteels et al., 1987). In such a way, indirect communication turns out to be particularly *well adapted to stable environments with relatively persistent sources of information*. For example, human trails are often strongly imprinted on the ground, which is suitable to shape urban green spaces, since entry and exit points barely evolve in time.

In contrast, **direct information transfer** tends to provide a higher reactivity to external changes and appears *more adapted to volatile information sources*. The consensus on the swimming direction adopted by fish schools is likely to suddenly change in response to the occurrence of novel information, such as a predator strike. Here, unlike indirect communication, information spreads directly from one individual to its neighbors, and the spatial proximity of the individuals allows the information to travel rapidly among them. In pedestrians, direct interactions allow people to optimize their movements in many regards, and lead to adapted collective answers to environmental perturbations such as obstacles or bottlenecks (Helbing et al., 2005). On the other hand, this higher flexibility often implies a lower level of selection of information, since weak random fluctuations can be amplified at the group level. In fish schools, for example, this may create useless movements that can be costly (Couzin, 2007). In general, the higher the interaction range, the less sensitive is the system to small perturbations, since information is locally integrated among a larger number of individuals.

**4.4 Self-Organized dynamics and individual complexity**

Throughout this paper, we relied on various human and animal systems to explore the mechanisms underlying the emergence of collective patterns. The described systems differ in many regards, and in particular in terms of cognitive abilities of the individuals. When investigating self-organization processes, however, it is common to reduce the level of complexity of group members to a set of simple behavioral rules. Therefore, the question of the relevance of this approach for sophisticated individuals (such as humans) arises. Moreover, which additional features can result from higher cognitive abilities at the level of the individual?

Obviously, the presence of common fundamental feedback mechanisms attests that some collective processes exhibited in human crowds can be explained without invoking complex decision-making abilities at the level of the individual. The success of simplified behavioral models in reproducing many emergent behaviors in crowds demonstrates that higher cognitive abilities are not *required* to capture the self-organized dynamics (Ball, 2004). In most cases, people react to well-known situations in a more or less automatic manner, promoting relatively predictable collective patterns similar to those produced in animal societies.

However, considering the wide variety of potential behavioral responses of complex beings, it is likely that individual complexity may play a role in the collective dynamics. *Individual learning* is a feature that can interfere with the collective dynamics. Human beings for instance, can quickly learn from past experiences, and adapt to new situations. As an illustration, we previously highlighted that pedestrian interactions may be biased by a side preference. This can be explained by considering the emergence of a behavioral convention, due to the ability of people to learn avoidance strategies from repeated interactions. As a result, what individuals learn affects the configuration of the emerging pattern. Since the learning process can be affected by numerous factors, behavioral conventions develop in different ways, depending on the geographical area: while Western European populations learned that avoidance on the right-hand side is preferable, some Asian countries similarly developed a left-hand preference.

Such learning processes play a role in animal societies as well, since many individual animals can also learn from their past experiences. Examples of learning involved in self-organized processes can be seen in the case of specialization of workers in insect societies. The more an individual performs a given task, the more it gets used to it and the faster it responds to this task in the future, leading to the emergence of specialized workers (Theraulaz et al., 1991, 1998; Ravary et al., 2007). Learning is not unique to human beings, but people are more prone to this kind of adaptation and new behavioral biases can evolve on shorter time scales, and for a larger variety of different settings. Interestingly, behavioral conventions are themselves self-enforcing and can spread across the population in a non-linear way, with no need of central authority (Helbing, 1992; Young, 1996). In terms of self-organized dynamics, such a learning process induces a common behavioral bias among individuals (by acting on the so-called spontaneous behavior, or on the behavioral response). Although weak, such a bias, affecting all individuals, is amplified through reinforcement loops, eventually resulting in a qualitative change of the collective response (see section 4.2).

## 5. Conclusion

In this contribution, we showed how a wide set of self-organized phenomena can be described and understood by means of local interaction mechanisms. Repeated interactions among individuals, random fluctuations, reinforcement loops and negative feedbacks are the basis of self-organization processes. The fact that a common approach can describe and explain the dynamics of various emerging collective behaviors strengthens the idea that these have a similar root, although the individuals involved differ in size, aims or cognitive capacities.

The discussion of various cases highlighted that individuals exchange information by means of direct or indirect interactions. This local exchange of information is then integrated at the collective level by means of feedback loops to produce adapted collective responses to various kinds of problems. Swarms and crowds consequently manage to take advantage of their numbers to cope with their complex environment and achieve sorting tasks, optimize their activities or reach consensual decisions. Furthermore, through learning processes, individuals can develop behavioral specificities that may have additional effects on the collective dynamics. In human societies, for example, the emergence of behavioral conventions can induce a common behavioral bias in the population that enhances in turn the self-organized dynamics.

## Acknowledgments

We thank the three reviewers and the editor for inspiring comments and discussions. Mehdi Moussaïd's doctoral fellowship is jointly financed by the ETH Zurich and the CNRS. Simon Garnier has a research grant from the French Ministry of Education, Research and Technology. MM and SG are grateful for partial financial support by grants from the CNRS (Concerted Action: Complex Systems in Human and Social Sciences) and the Université Paul Sabatier (Aides Ponctuelles de Coopération).

# Figures & Table

Table 1: Summary of case studies

| SYSTEM | SPONTANEOUS BEHAVIOR | RELATED INFORMATION | BEHAVIORAL RESPONSE | INFORMATION SUPPORT |
|---|---|---|---|---|
| People looking up (Milgram experiment) | Weak probability to look up | "Direction of a point of interest" | - Increased probability to look up<br>- Weighted by the number of people looking up | - Direct information transfer<br>- Visual signals |
| Digg.com | Read random stories | "Interesting news" | - Increased probability to read the news<br>- Weighted by the number of diggs | - Indirect information transfer<br>- Virtual signals (diggs) |
| Foraging ant trails | Random walk<br>Biased by environment (e.g. borders, walls) | "Location of a food source" | - Attraction along the pheromone trail<br>-Weighted by concentration of pheromone | - Indirect information transfer<br>- Chemical signals (pheromones) |
| Pedestrians trails | Goal-oriented motion<br>Biased by environment (attractive places) | "Short and comfortable path" | - Attraction toward the trail<br>- Weighted by trail visibility | - Indirect information transfer<br>- Physical signals (alteration of the ground) |
| Fish schooling | Turns randomly<br>Potentially biased toward attractive places (food source, migration route) | "Moving direction" | - Move in the average perceived direction. | - Direct information transfer<br>- Visual signals combined with water displacement |
| Clapping synchronization | Clap at own rhythm | "Clapping rhythm" | - Adjust clapping to perceived average | - Direct information transfer<br>- Acoustic signals |
| Lane formation in ants | Goal-oriented motion along a pheromone trail | "Faster moving area" | - Change moving direction<br>- Weighted by amount of load | - Direct information transfer<br>- Physical contacts |
| Lane formation in pedestrians | Goal-oriented motion | "Faster and more comfortable walking area" | - Move away from perceived people | - Direct information transfer<br>- Visual signals |

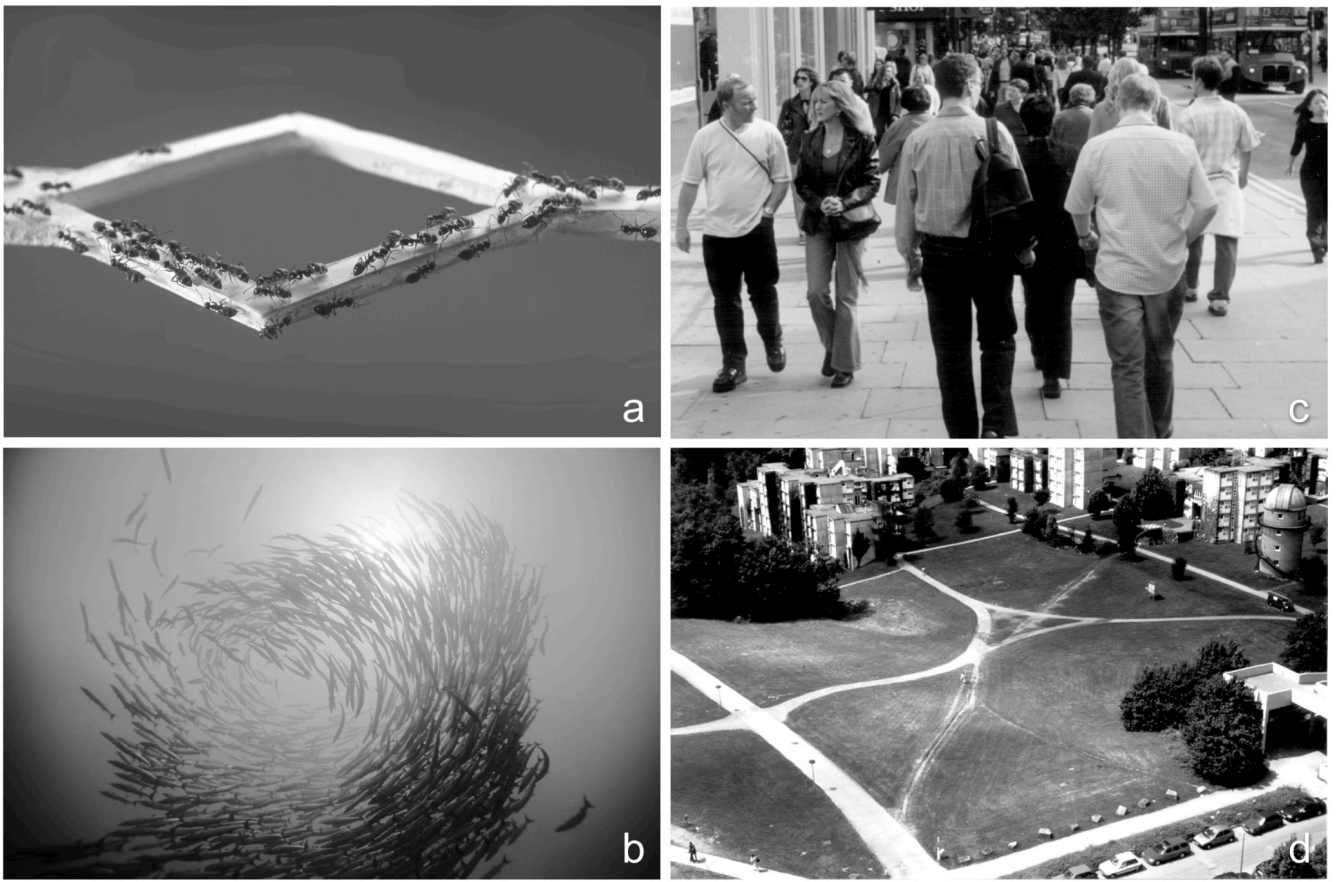

Figure 1: Examples of self-organized phenomena in human and animal populations.

a) Trail formation and collective path selection in ants. The figure refers to an experiment with a two-paths-bridge linking the nest and a food source. b) Emergence of a vortex in a school of fish, consisting of individuals circling around an unoccupied core (© Tammy Peluso, istockphoto.com) c) Segregation of a bidirectional flow of pedestrians into lanes of people with a common walking direction (from Helbing et al., 2005) d) Human trails formed on the University campus of Stuttgart-Vaihingen (from Helbing et al., 1997).

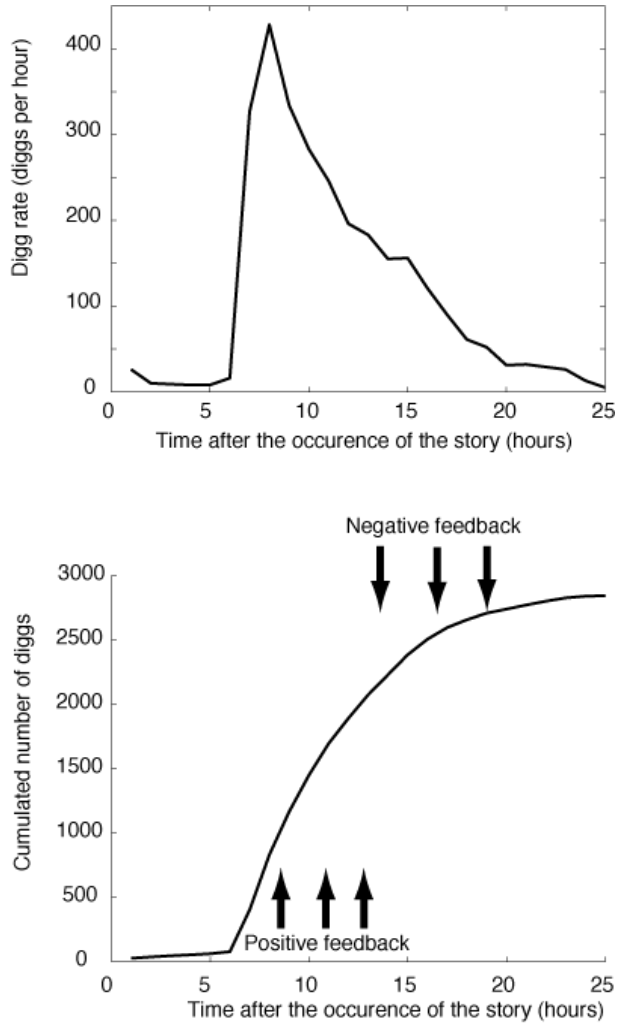

Figure 2: Observed dynamics for a story on *digg.com* during one day
*Top*: Observed digg-rate for a given story. The sudden amplification of interest after 5 hours is due to the reinforcement effect of the increased the number of diggs, while the following decay results from the decreasing attention of users. *Bottom*: Cumulative number of diggs illustrating the antagonist effects of positive and negative feedbacks (same dataset).

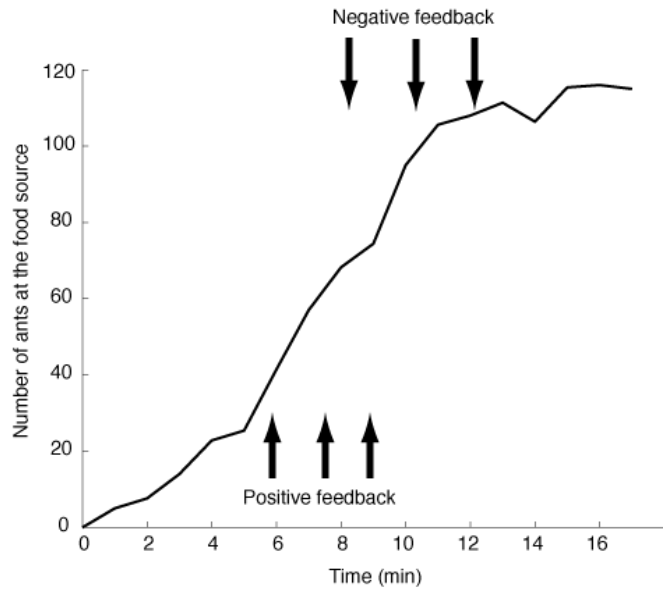

Figure 3: Recruitment dynamics in the ant *Linepithema humile*.
Observation of number of ants involved in a foraging task, illustrating the emergence of a pheromone trail between the nest and a food source (unpublished experimental data). While an increasing pheromone concentration attracts more and more ants along the trail during the first moments, the jamming that occurs around the food source at higher density counterbalances the previous amplification and stabilizes the flow of ants at a constant level.

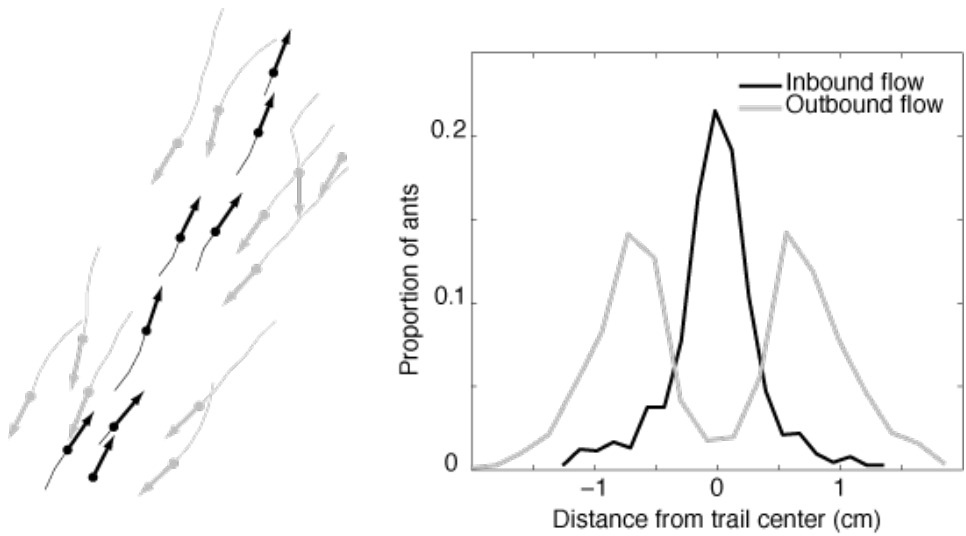

Figure 4: Lane formation in a simulation of bidirectional traffic of army ants *Eciton burchelli*
*Left*: Snapshot of simulation (after Couzin and Franks, 2002). The dark arrows represent ants loaded with prey and going back to the nest, while light arrows represent ants leaving the nest. *Right*: Distribution of ants of the two flows with respect to the trail center, illustrating the spatial segregation of inbound and outbound ants.

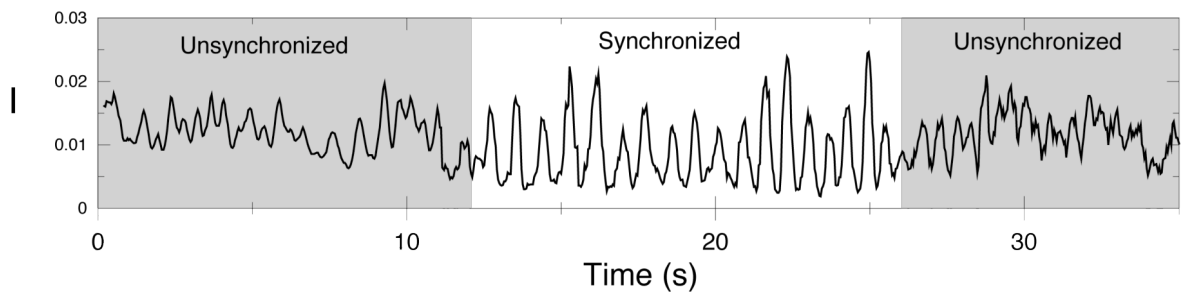

Figure 5: Acoustic signal of a clapping audience recorded after a theater performance in Hungary. The typical pattern consists in an alternation of synchronized and unsynchronized applause phases (after Nèda et al., 2000).

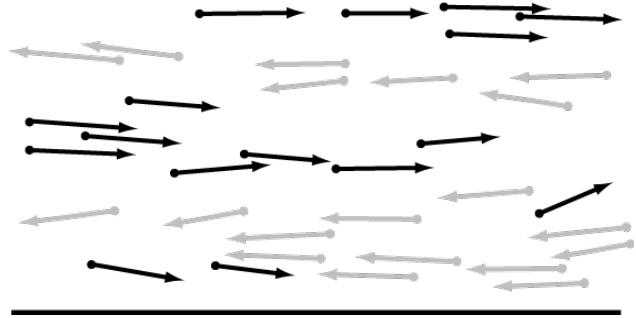

Figure 6: Lane formation in pedestrians
Snapshot of a simulation of bidirectional flows of pedestrians, reproducing the spontaneous emergence of lanes (after Helbing and Molnar, 1995).